\documentclass{article}
\usepackage{spconf,amsmath,graphicx}
\usepackage{algorithm} 
\usepackage{algpseudocode}  
\usepackage{amsmath}  
\usepackage{caption}
\usepackage{subfigure}
\usepackage{multirow}
\usepackage{stfloats}
\usepackage{url}
\usepackage{caption}
\title{END-TO-END LEARNED IMAGE COMPRESSION WITH FIXED POINT WEIGHT QUANTIZATION}
\name{Heming Sun$^\ast$$^\dagger$, Zhengxue Cheng$^\ddagger$, Masaru Takeuchi$^\ast$, Jiro Katto$^\ast$$^\ddagger$\thanks{This work was supported in part by JST, PRESTO Grant Number JPMJPR19M5, Japan, and in part by Hoso Bunka Foundation.}}
\address{$^\ast$ Waseda Research Institute for Science and Engineering, Waseda University, Tokyo, Japan \\
$^\dagger$ JST, PRESTO, 4-1-8 Honcho, Kawaguchi, Saitama, Japan \\
$^\ddagger$ Department of Computer Science and Communication Engineering, Waseda University, Tokyo, Japan
}
\begin{document}	
	
%
\maketitle
\begin{abstract}
Learned image compression (LIC) has reached the traditional hand-crafted methods such as JPEG2000 and BPG in terms of the coding gain. However, the large model size of the network prohibits the usage of LIC on resource-limited embedded systems. This paper presents a LIC with 8-bit fixed-point weights. First, we quantize the weights in groups and propose a non-linear memory-free codebook. Second, we explore the optimal grouping and quantization scheme. Finally, we develop a novel weight clipping fine tuning scheme. Experimental results illustrate that the coding loss caused by the quantization is small, while around 75\% model size can be reduced compared with the 32-bit floating-point anchor. As far as we know, this is the first work to explore and evaluate the LIC fully with fixed-point weights, and our proposed quantized LIC is able to outperform BPG in terms of MS-SSIM.
\end{abstract}
\section{Introduction}
\label{sec:intro}

Image compression is important to relieve the burden of the image transmission and storage. In the past decades, several standards have been developed such as JPEG \cite{wallace1992jpeg}, JPEG2000 \cite{rabbani2002overview}, WebP \cite{lian2012webp} and HEVC intra (BPG) \cite{sullivan2012overview} . Different from the hand-crafted ways, deep learning has shown a promising compression ability as reported in \cite{rippel2017real}, \cite{toderici2017full},\cite{balle2018variational},\cite{cheng2018deep},\cite{cheng2019learning},\cite{cheng2019deep}. By employing a proper neural network structure and enhanced probability models such as factorized and hyper prior, learned image compression (LIC) has outperformed the BPG in terms of MS-SSIM. Though LIC methods can achieve a good coding gain, utilizing many layers and channels will enlarge the network model, which prohibits the potential usage of LIC on resource-limited embedded devices.

Recently, weight quantization has shown a superior capability for the model compression in many networks such as AlexNet, ResNet and GoogleNet. A binary network with 1-bit weight and activation were proposed in \cite{bnn} . Similarly, the number of bits can be reduced to two in a ternary weight network \cite{ternery}. \cite{han2015deep} pruned the network and quantized each network connection from 32-bit to 5-bit. An incremental quantization scheme with fine tuning was proposed in \cite{zhou2017incremental}. \cite{park2019cell} kept the network accuracy while dividing the weights to the arbitrary bit-widths. \cite{gong2014compressing} exploited the vector quantization to achieve 16-24 times compression ratio. An optimized quantization scale is learned in \cite{choi2018pact} to achieve a 4-bit precision at a comparable accuracy with full precision models.

Though quantization has achieved good performance for many popular models, there is almost no related work for LIC. \cite{balle2018integer} designed an integer network and provided a heuristic training scheme. However, using integer networks for main path will diminish the coding gain. In this paper, we quantize the weights to 8-bit fixed-point for both main path and hyper path by 1) formulating the quantization in grouping, 2) determining the optimal grouping and quantization scheme by coding gain, and 3) proposing a weight clipping fine tuning method. As a result, we can reduce about 75\% model size compared with the 32-bit floating-point edition, and outperform BPG in terms of MS-SSIM.

\section{Quantization Method for LIC}
\label{sec:format}

\subsection{Formulation of Baseline Hyperprior Architecture}
\label{ssec:subhead}

\begin{figure}[t]
	\begin{minipage}[b]{1.0\linewidth}
		\centering
		\centerline{\includegraphics[width=7.0cm]{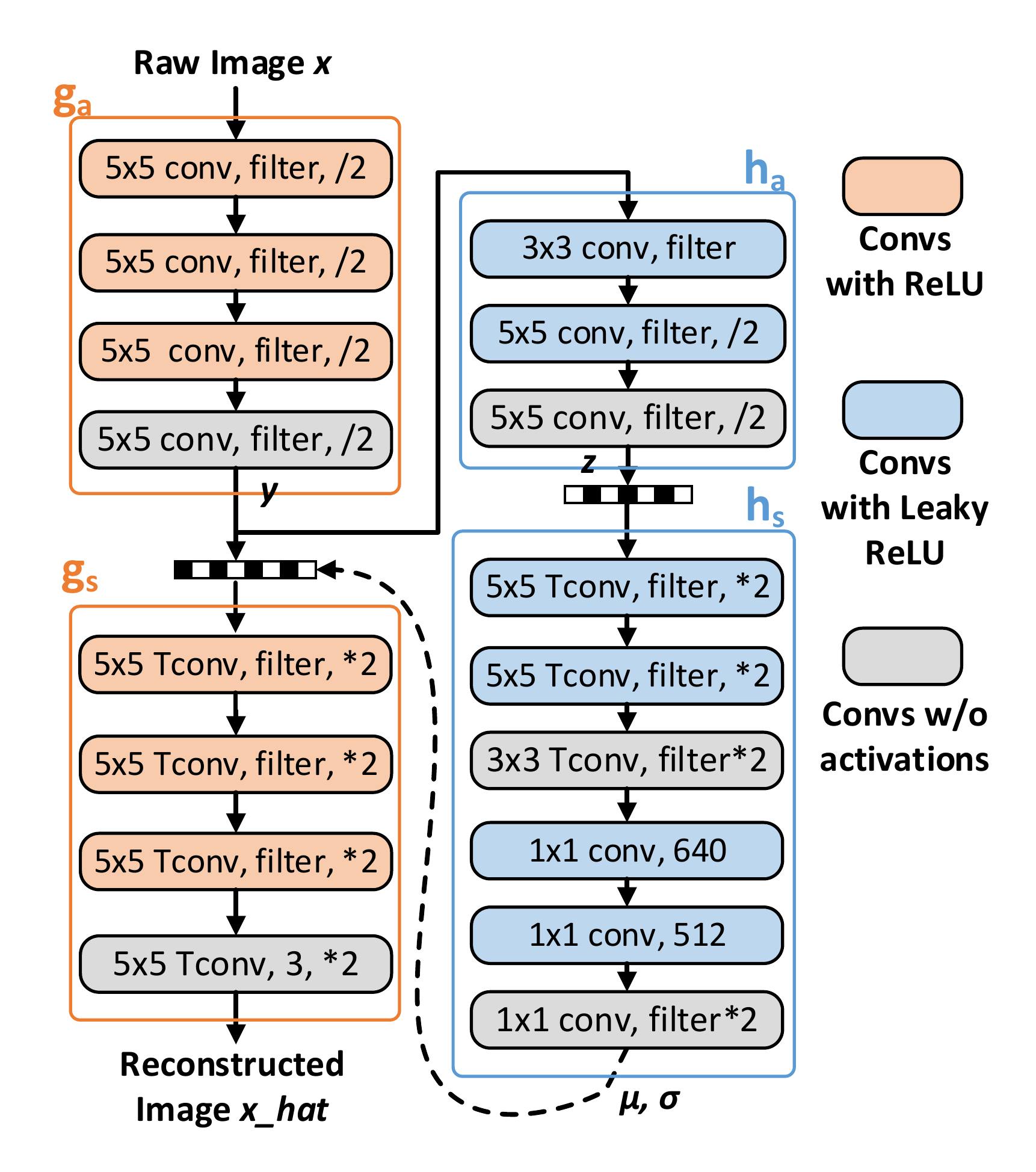}}
	\end{minipage}
	\caption{Diagram of mean-scale hyperprior network.}
	\label{fig_hyper}
\end{figure}

The baseline network we used is shown in Fig. \ref{fig_hyper}, which is the \textit{hyperprior-5} in \cite{cheng2019deep}. The operation can be formulated as the following two equations
\begin{equation}
\small
\begin{split}
	& y=g_a(x;\phi) \\
	& \hat{y}=Q(y) \\
	& \hat{x}=g_s(\hat{y};\theta) \\
\end{split}
\label{eq_main}
\end{equation}
\begin{equation}
\small
\begin{split}
& z=h_a(y;\phi_h) \\
& \hat{z}=Q(z) \\
& \mu_y,\sigma_y=h_s(\hat{z};\theta_h)
\end{split}
\label{eq_hyper}
\end{equation}
\noindent where Eq. \ref{eq_main} and Eq. \ref{eq_hyper} defines the main and hyper path operation respectively. $x$ and $\hat{x}$ are the raw and reconstructed image, $y$ and $z$ are two-layer latent nodes that will become $\hat{y}$ and $\hat{z}$ through a uniform quantization. $\phi$ and $\theta$ are the trained parameters. $\mu_y$ and $\sigma_y$ are the estimated mean and variance for the usage of the probability model of $y$.

About the activation function, all the layers in $g_a$ and $g_s$ utilized ReLU, while all the layers in $h_a$ and $h_s$ used leaky-ReLU. Noted that there is no activations for the final layer in the analysis and synthesis transforms.

\subsection{Grouping and Quantization Formulation}
\label{ssec:subhead}

\begin{figure}[t]
	\small
	\begin{minipage}[b]{1.0\linewidth}
		\centering
		\centerline{\includegraphics[width=8.0cm]{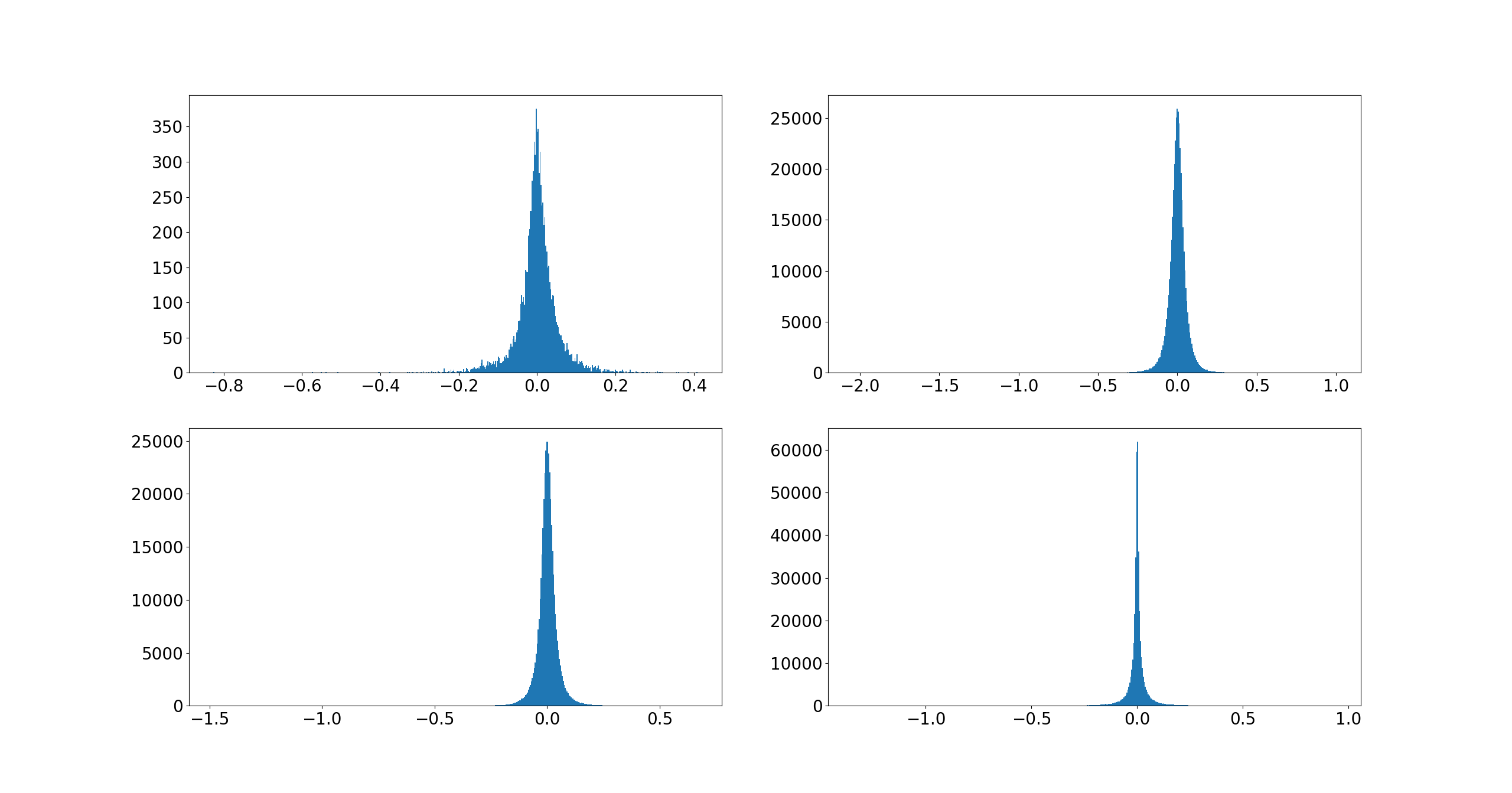}}
		\centerline{(a) Four layers of analysis transform $g_a$.}\medskip
	\end{minipage}
	\begin{minipage}[b]{1.0\linewidth}
		\centering
		\centerline{\includegraphics[width=8.0cm]{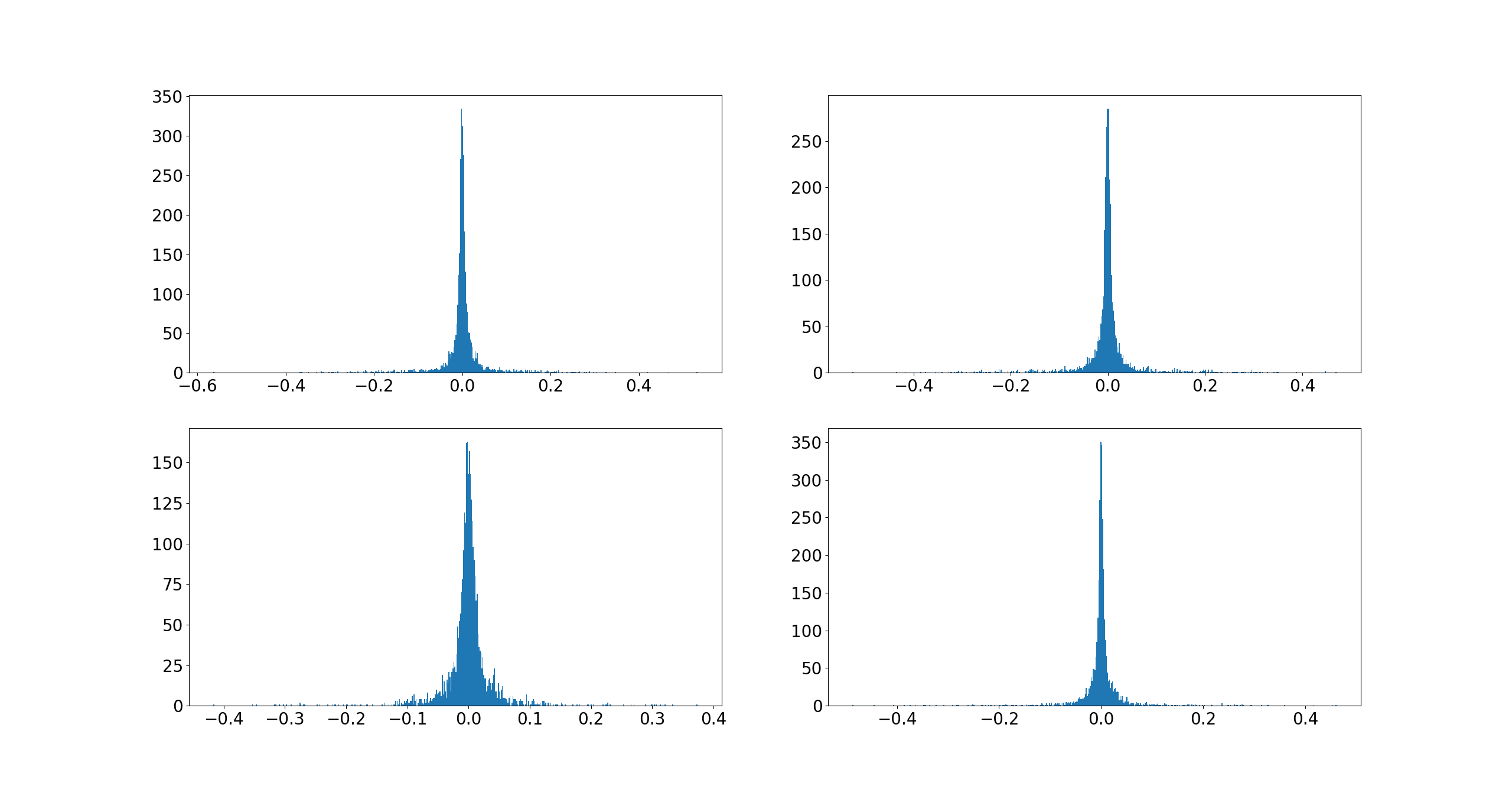}}
		\centerline{(b) First four channels of the final layer of analysis transform $g_a$.}\medskip
	\end{minipage}
	\caption{Histograms of layer-wise and channel-wise weights.}
	\label{fig_weight}
\end{figure}

At first, we visualize the weight histograms in Fig. \ref{fig_weight}. We can see that both layer-wise and channel-wise weights follow the Gaussian distribution well with zero mean. According to \cite{isscc}, 8-bit integer can save around 19x multiply and 30x accumulation  power reduction compared with the 32-bit floating point. Moreover, 8-bit can align with the bit-width of most on-chip memories. Therefore, the bit budget is set as 8-bit in this work, which can be represented as a fixed-point manner \{1,$IL$,$FL$\} where 1 is the signed bit, $IL$ and $FL$ stands for the integer and fractional bits.

First, we scale the weights $\textbf{w}{\in}R$ to \textbf{sw}$\in$(-2, 2), and then set $IL$ and $FL$ as 1-bit and 6-bit, respectively. By doing so, the most significant bit for $IL$ and $FL$ will not be wasted. The scaling is performed in groups. For the weights in the k-th group, $\textbf{w}_k$ are scaled with a scalar scaling factor $(\textit{sf}_k)$.
\begin{equation}
\small
\textbf{sw}_k=\textbf{w}_k \times \textit{sf}_k
\label{eq_3}
\end{equation}
\noindent where $\textit{sf}_k$ can be calculated by Eq. \ref{eq_4} which can be easily implemented as a shift operation in the hardware.
\begin{equation}
\small
\textit{sf}_k =   2 ^ { - \lfloor \log _2{ max ( | \textbf{w}_k | ) } \rfloor  } 
\label{eq_4}
\end{equation}
\indent After scaling, for each scaled weight element $sw_i$ in the $k$-th group, the quantization is conducted as follows
\begin{equation}
\small
Q(sw_i^k)=q_j^k
\label{eq_5}
\end{equation}
\noindent where $j\in 0,...,2^{FL}-1$. For the linear quantization (LQ), the operation of each group is performed as follows
\begin{equation}
\small
q_j=  \dfrac{\lfloor sw_i\times2^{FL} \rfloor }{2^{FL}} +\dfrac{\xi_i}{2^{FL}}
\label{eq_lq}
\end{equation}
\noindent where $\xi_i$ is the rounding function in Eq. \ref{eq_7}.
\begin{equation}
\small
\xi_i=\begin{cases}
1, \text{  if } sw_i\times2^{FL}-\lfloor sw_i\times2^{FL} \rfloor > 0.5\\
0, \text{  otherwise}
\end{cases}
\label{eq_7}
\end{equation}
\indent For the non-linear quantization (NLQ), each $q_j$ is determined according to the distributions of weights. One method is to use Lloyd's method \cite{lloyd1982least}, so that the codebook (i.e. $q_j$) can be optimized according to the following equation
\begin{equation}
\small
min\sum_{i=0}^{ len(\textbf{w}_k)-1} \sum_{j=0}^{2^{FL}-1} ||sw_i-q_j||_2^2
\label{eq_lloyd}
\end{equation}
\indent Despite LLoyd's algorithm is optimal, it requires hardware cost such as Look Up Table (LUT) to memorize the codebook for each group. To relieve the memory overhead, we developed an alternative memory-free codebook as shown in Eq. \ref{eq_nlq}. By doing so, we only need to compare $sw_i$ with power of two (i.e. 0.25 and 0.5) to obtain the quantized result $q_j$ at runtime so that there will be no memory consumption for the codebook.
\begin{equation}
\small
q_j = \begin{cases}
 \dfrac{ \lfloor sw_i\times2^{FL-1} \rfloor  }{2^{FL-1}} +\dfrac{\xi_i}{2^{FL-1}}, & \text{if } |sw_i|\in[0.5,2)\\
\dfrac{ \lfloor sw_i\times2^{FL}  \rfloor }{2^{FL}} +\dfrac{\xi_i}{2^{FL}}, & \text{if } |sw_i|\in[0.25,0.5)\\
\dfrac{ \lfloor sw_i\times2^{FL+2} \rfloor }{2^{FL+2}} +\dfrac{\xi_i}{2^{FL+2}}, & \text{else} \\
\end{cases}
\label{eq_nlq}
\end{equation}

\subsection{Quantization and Grouping Scheme Determination}
\label{ssec:subhead}

As described in the above, scaling is conducted in groups so that scaling factor ($\textit{sf}$)  for each group has to be stored. With more groups, there will be more non-zero weights after the quantization, while more consumption is required to store $\textit{sf}$. In this paper, we explore two structured group scheme that is layer-wise (LW) and channel-wise (CW) grouping.

In the case of LW grouping, each group contains the weight $\textbf{w} \in R^4$ where four dimension represents input channel, kernel width, kernel height and output channel. In the case of CW grouping, each group contains the weight $\textbf{w} \in R^3$ where three dimension represents input channel, kernel width and height. For both grouping scheme, we attempt the method in Eq. \ref{eq_lq} and Eq. \ref{eq_nlq}. Besides, we also exploit Eq. \ref{eq_lloyd} for LW rather than CW since it is not feasible to memorize LUTs for all the channels. For the model with a moderate rate ($\lambda$=0.015, MSE optimized, 4x$10^5$ iterations), the results are shown in Table \ref{table_1}. From the results, we can conclude that using CW-NLQ can reach the best coding gain.

\begin{table}[t]
	\setlength{\abovecaptionskip}{0.0cm}   
	\setlength{\belowcaptionskip}{0.0cm}
	\caption{Comparison of different quantization and grouping in terms of coding gain}
	\footnotesize 
	\begin{center}
	{
		\begin{tabular}{|c|c|c|c|c|c|c|}
			\hline
			
			\textbf{ }					&\text{ }			&\multicolumn{3}{|c|}{\text{Layer-wise}}			   		 &\multicolumn{2}{|c|}{\text{Channel-wise}}				\\ \hline
			\textbf{ }					&\text{Origin}		&\text{LQ}			&\text{NLQ}			&\text{Lloyd}		&\text{LQ} 			&\text{NLQ}			\\ \hline
			
			\textbf{PSNR (dB)}			&\text{$32.32$}		&\text{$28.81$}		&\text{$32.08$}		&\text{$32.18$}		&\text{$31.60$} 		&\text{$32.25$}		\\ \hline
			
			\textbf{bpp}				&\text{$0.529$}		&\text{$0.770$}		&\text{$0.555$}		&\text{$0.537$}			&\text{$0.569$} 		&\text{$0.539$}		\\ \hline
			
		\end{tabular}
	}
	\end{center}
	\label{table_1}
	\vspace{-6mm}
\end{table}

For the NLQ in Eq. \ref{eq_nlq}, the least magnitude that will not be quantized to zero is $\frac{1}{2^{FL+2}}$. To ensure this value is enough for the precision, we explore the coding gain of CW-LQ with different precisions as shown in Table \ref{table 2}. We can see that when increasing the precision from $\frac{1}{2^{FL}}$ to $\frac{1}{2^{FL+2}}$, there are obvious improvements for both PSNR and bpp. However, when further increasing the precision to $\frac{1}{2^{FL+3}}$, there is only 0.02dB and 0.002bpp difference that is quite trivial. Therefore, we decide to use $\frac{1}{2^{FL+2}}$ as the highest precision.

\begin{table}[t]
	\setlength{\abovecaptionskip}{0.0cm}   
	\setlength{\belowcaptionskip}{0.0cm}
	\caption{Comparison of different precision in terms of coding gain}
	\footnotesize 
	\begin{center}
		{
			\begin{tabular}{|c|c|c|c|c|c|c|}
				\hline
				\textbf{Precision}					&\text{$\frac{1}{2^{FL}}$}		&\text{$\frac{1}{2^{FL+1}}$}			&\text{$\frac{1}{2^{FL+2}}$}		&\text{$\frac{1}{2^{FL+3}}$}			\\ \hline
				
				\textbf{PSNR (dB)}			&\text{$31.60$}		&\text{$32.17$}		&\text{$32.28$}		&\text{$32.30$}		\\ \hline
				
				\textbf{bpp}				&\text{$0.569$}		&\text{$0.541$}		&\text{$0.532$}		&\text{$0.530$}			\\ \hline
				
			\end{tabular}
		}
	\end{center}
	\label{table 2}
	\vspace{-6mm}
\end{table}

\subsection{Weight Clipping Fine Tuning (WCFT)}
\label{ssec:subhead}

By using proposed CW-NLQ, the coding loss has been trivial as shown in Table \ref{table_1}. However, according to the experimental results, for the higher rate models, the loss of CW-NLQ is still not negligible.

As described in the above, the least magnitude of $|sw_i|$ that will not be quantized to zero is $\frac{1}{2^{FL+2}}$. Therefore, to generate more non-zero quantized results, larger $|sw_i|$ is desired. From Eq. \ref{eq_4}, we can see that $\textit{sf}_k$ is fully dependent on $max(|\textbf{w}_k|)$, and $\textit{sf}_k$ will become larger with a smaller $max(|\textbf{w}_k|)$. Therefore, our target is to reduce $max(|\textbf{w}_k|)$.

First, we train the network as usual to obtain the optimized $\textbf{w}$. After that, for each group, we clip the maximum magnitude to $2^{\lfloor \log _2{ max ( | \textbf{w}_k | ) } \rfloor} - \epsilon$ where $\epsilon$ is a very small value. After the clipping, we fine tune the network to compensate the loss caused by the clipping. During the fine tuning, we adopt the straight-through estimator \cite{hinton2012neural} that is to preserve the gradient and cancels the gradient when $w_i$ is larger than the clipping value as shown in Eq. \ref{eq_ste}. The pseudo code of overall training procedures with fine tuning is given in \textbf{Algorithm 1}.
\begin{algorithm} [t]
	\caption{\small Proposed Training Method with Weight Clipping Fine Tuning}
	\label{alg:Framwork}
	\begin{algorithmic}[1]
		\small
		\For{ number of training iterations \textit{$I_1$} }
		
		\State \emph{ J($g_a, g_s, h_a, h_s$) = $\lambda$  D(x,$\hat{x}$)+R($\hat{y}$) + R($\hat{z}$)  } 
		\State Update $|\textbf{w}_k|$ of \emph{$g_a, g_s, h_a, h_s$} by descending its SGD
		\EndFor
		
		\State Clip $|\textbf{w}_k|$ to $ 2^  {  \lfloor \log _2{ max ( | \textbf{w}_k | ) } \rfloor  }  - \epsilon $
		
		\For{ number of fine tuning iterations \textit{$I_2$} }
		\State \emph{ J($g_a, g_s, h_a, h_s$) = $\lambda$  D(x,$\hat{x}$)+R($\hat{y}$) + R($\hat{z}$)  } 
		\State Update $|\textbf{w}_k|$ with straight-through estimator (STE)
		\EndFor
	\end{algorithmic}
\end{algorithm}
\begin{equation}
\small
w'_i=\begin{cases}
w_i, &\text{  if } w_i \leq 2^{\lfloor \log_2{ max ( | \textbf{w}_k | ) } \rfloor} - \epsilon\\
2^{\lfloor \log_2{ max ( | \textbf{w}_k | ) } \rfloor} - \epsilon, &\text{  otherwise}
\end{cases}
\label{eq_clip}
\end{equation}
\begin{equation}
\small
g_{w_i} =  g_{w'_i}1_{|w_i|\leq 2^{\lfloor \log_2{ max ( | \textbf{w}_k | ) } \rfloor} - \epsilon }
\label{eq_ste}
\end{equation}
\section{Experimental Results}
\label{sec:pagestyle}

\subsection{Network and Training Details}
\label{ssec:subhead}

For the training, we use 256$\times$256 patches cropped from ImageNet \cite{deng2009imagenet}, and set batch size as eight. $I_1$ and $I_2$ in \textbf{Algorithm 1} are set as $10^6$ and $10^5$. The loss function is given in the following equation
\begin{equation}
\small
\emph{ J = $\lambda$  D(x,$\hat{x}$)+R($\hat{y}$) + R($\hat{z}$)  } 
\label{eq 2.5}
\end{equation}
\noindent where $D(x,\hat{x})$ is MSE and MS-SSIM to optimize PSNR and MS-SSIM, respectively, $R(\hat{y})$ and  $R(\hat{z})$ are the consumed bits of $y$ and $z$. $\lambda$ is set as [0.001625, 0.00325, 0.0075, 0.015, 0.03, 0.05] for the MSE, and [3,5,10,40,80,128] for the MS-SSIM to generate six models, respectively. For both MSE and MS-SSIM, we use 128 filters for four lower rate models, and 192 filters for two higher rate models. We adopt the WCFT for the two higher rate models.

\subsection{Coding Performance Evaluation}
\label{ssec:subhead}

\begin{figure}[t]
	\begin{minipage}[b]{1.0\linewidth}
		\centering
		\centerline{\includegraphics[width=7.5cm]{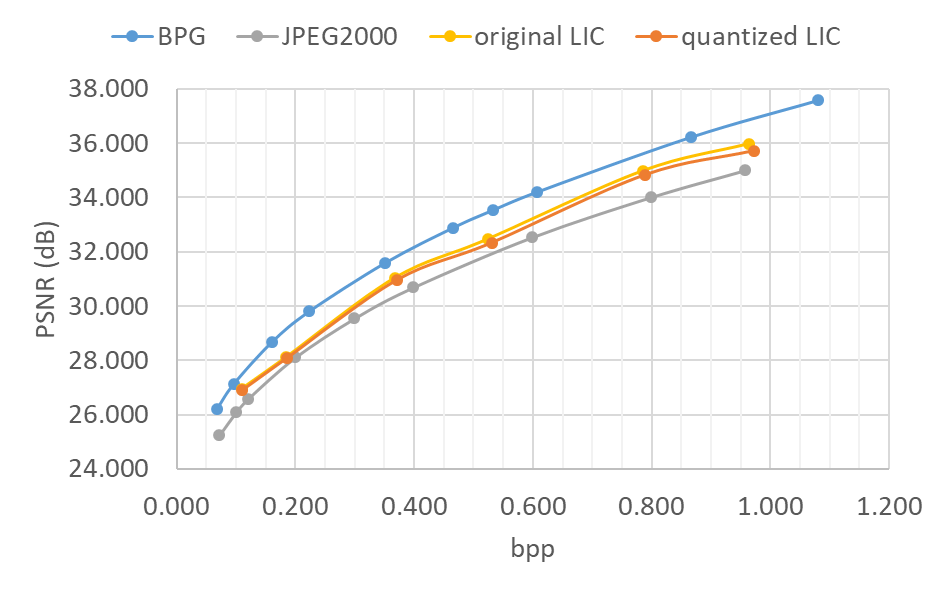}}
	\end{minipage}
	\caption{PSNR comparison of 32-bit floating-point original LIC, proposed 8-bit fixed-point quantized LIC, BPG and JPEG2000.}
	\label{fig_psnr}
\end{figure}

\begin{figure}[t]
	\begin{minipage}[b]{1.0\linewidth}
		\centering
		\centerline{\includegraphics[width=7.5cm]{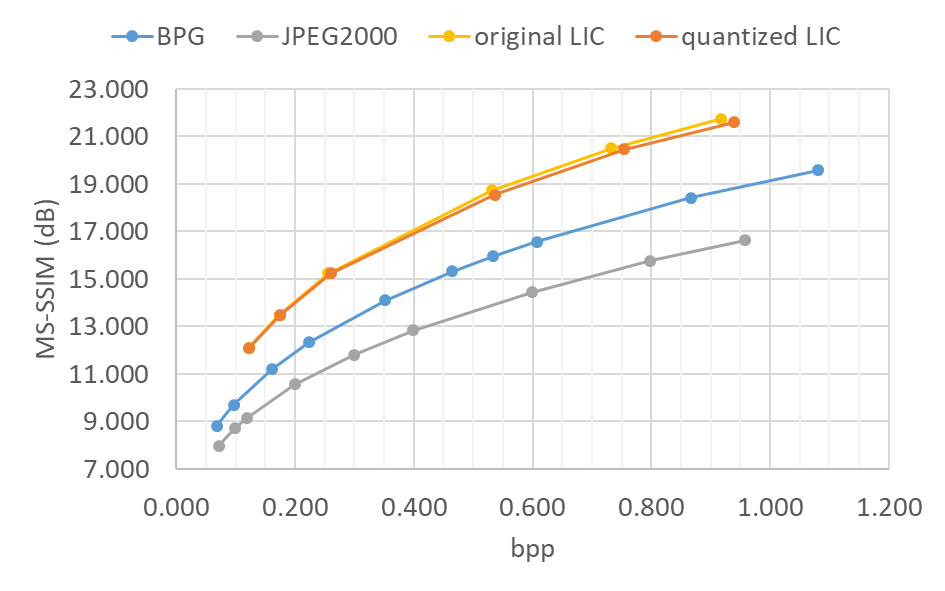}}
	\end{minipage}
	\caption{MS-SSIM comparison of 32-bit floating-point original LIC, proposed 8-bit fixed-point quantized LIC, BPG and JPEG2000.}
	\label{fig_ssim}
\end{figure}

First, we evaluate the coding gain of our proposal by the Kodak dataset \cite{kodak} with 24 distortion-free images, and the results are shown in Fig. \ref{fig_psnr} and Fig. \ref{fig_ssim}. We can see that our proposed 8-bit fixed-point LIC is quite close to original 32-bit floating-point edition. For the four middle rate models, the BD-psnr \cite{bjontegaard2001calculation} loss compared with the original anchor is only 0.1183dB and 0.1486dB for MSE and MS-SSIM, respectively. Besides, we can outperform JPEG2000 in terms of PSNR and perform better than BPG in terms of MS-SSIM. Noted that the PSNR of fixed-point LIC can be further improved by enhancing the floating-point anchor model.

\begin{table}[t]
	\setlength{\abovecaptionskip}{0.0cm}   
	\setlength{\belowcaptionskip}{0.0cm}
	\caption{Coding gain improvement by using proposed WCFT}
	\scriptsize
	\begin{center}
		{
			\begin{tabular}{|c|c|c|c|c|c|c|} \hline
				
				\textbf{}  &{\text{ }} 	&\multicolumn{2}{|c|}{\text{MSE}}&\multicolumn{2}{|c|}{\text{MS-SSIM}}  	\\ \hline		
				
				\textbf{}  &$\lambda$ 			&\text{$0.03$}		&\text{$0.05$}			&\text{$80$}		&\text{$128$}				\\ \hline
				
				\multirow{2}*{\textbf{w/o WCFT}} 	&\text{PSNR (dB)}		&\text{$-0.382$}		&\text{$-0.334$}		&\text{$-0.285$}		&\text{$-0.509$}		\\\cline{2-6}
				
				\textbf{} &\text{bpp}&\text{$+0.019$}		&\text{$+0.012$}		&\text{$+0.018$}		&\text{$+0.038$}				\\ \hline
				
				\multirow{2}*{\textbf{with WCFT}}	&\text{PSNR (dB)}		&\textbf{$-0.148$}		&\textbf{$-0.258$}		&\textbf{$-0.107$}		&\textbf{$-0.264$} \\ \cline{2-6}
				
				\textbf{} &\text{bpp}&\textbf{$+0.004$}		&\textbf{$+0.009$}		&\textbf{$+0.015$}		&\textbf{$+0.013$}		\\ \hline
				
			\end{tabular}
		}
	\end{center}
	\label{table_bdq}
	\vspace{-6mm}
\end{table}

We also evaluate the effect of the proposed WCFT in Table \ref{table_bdq}. Before using this scheme, for two high rate MSE models, the coding loss caused by the quantization is 0.382dB and 0.334dB, while it can be reduced to 0.148dB and 0.258dB. In addition, the bit increment caused by the quantization can also be reduced. Without using WCFT, the bpp is increased by 0.019 and 0.012, while the bpp is only increased by 0.004 and 0.009 after using WCFT. For two high rate MS-SSIM models, the coding loss can be decreased from 0.285dB and 0.509dB to 0.107dB and 0.264dB, respectively. The bpp increment can be reduced from 0.018 and 0.038 to 0.015 and 0.013, respectively. Therefore, using WCFT is quite helpful for the coding gain improvement.

\subsection{Memory Consumption Evaluation}
\label{ssec:subhead}

\begin{table}[t]
	\setlength{\abovecaptionskip}{0.0cm}   
	\setlength{\belowcaptionskip}{0.0cm}
	\caption{LIC model size comparison (MB)}
	\scriptsize
	\begin{center}
		{
			
			\resizebox{0.95\width}{!}{
			
			\begin{tabular}{|c|c|c|c|c|c|c|}
				\hline
				\textbf{Filter}							&\multicolumn{3}{|c|}{\text{$128$}}			&\multicolumn{3}{|c|}{\text{$192$}}			\\ \hline
				
				\textbf{Component}						&\text{weight}		&\textit{sf}	&\text{total}	&\text{weight}		&\textit{sf}	&\text{total}	\\ \hline
				
				\textbf{Original}			&\text{$20.72$}		&\text{$0$}	&\text{$20.72$}	&\text{$44.04$}	&\text{$0$}	&\text{$44.04$}	\\ \hline
				
				\textbf{Proposed}			&\text{$5.18$}		&\text{$0.0016$}	&\text{$5.1816$}	&\text{$11.01$}	&\text{$0.0021$}	&\text{$11.0121$}	\\ \hline
				
			\end{tabular}
		
			}
		
		}
	\end{center}
	\label{table_model}
	\vspace{-6mm}
\end{table}
We evaluate the model size comparison in Table \ref{table_model}. For the network structure in Fig. \ref{fig_hyper}, the number of bytes for the weight can be calculated by Eq. \ref{eq_original_weight} where $I$, $O$, $H$, $W$ are input channel number, output channel number, kernel height and width, and $i$ is the index of convolution layers. Overall, we have 17 layers. After quantizing each weight to 8-bit, the total weight storage can become one-fourth while there is memory overhead to store $\textit{sf}$. According to our experiments, 4-bit is adequate to save one scalar $\textit{sf}$. In the case of CW grouping, the number of $\textit{sf}$ is equal to the number of output channels. Overall, required bytes for the weights can be obtained by Eq. \ref{eq_proposed_weight}. From the results, we can see that about 75\% memory consumption can be saved since the overhead of the additional scaling factor is negligible compared with the storage of weight itself.
\begin{equation}
\small
\textbf{Original model size}
=\sum_{i=0}^{layer-1} I_i \times H_i  \times W_i \times O_i \times 4
\label{eq_original_weight}
\end{equation}
\begin{equation}
\small
\begin{aligned}
\textbf{Proposed model size} &=\sum_{i=0}^{layer-1} I_i \times H_i  \times W_i \times O_i \times 1 \\
&+\sum_{i=0}^{layer-1}O_i \times 0.5
\end{aligned}
\label{eq_proposed_weight}
\end{equation}

\section{Conclusions}
\label{sec:typestyle}

This paper proposes a fixed-point weight quantization method for LIC. First, we explore different kinds of grouping and quantization schemes, and then determine the optimal one based on the coding gain. In addition, to alleviate the coding performance loss caused by the quantization error, a fine tuning method is proposed. The results show that we can outperform the BPG in terms of MS-SSIM. For the future work, we will quantize the activations by fixed-point arithmetic and design the corresponding hardware architectures such as FPGA and ASIC.


\bibliographystyle{IEEEbib}
\bibliography{strings,refs}

\end{document}